\documentclass[a4paper,10pt]{article}
\usepackage[latin1]{inputenc}
\usepackage{graphics}
\usepackage{url}
\usepackage{cite}
\newcommand{\ED}[1]
{} 
\begin{document}
\title{Benchmarking some Portuguese S\&T system research units: 2nd Edition}
\author{Francisco M. Couto \and Daniel Faria \and Bruno Tavares \and Pedro Gonçalves \and Paulo Verissimo \and \\ %
\small LaSIGE, University of Lisbon, Portugal %
}
\date{}
\maketitle
\begin{abstract}
The increasing use of productivity and impact metrics for evaluation
and comparison, not only of individual researchers but also of
institutions, universities and even countries, has prompted the
development of bibliometrics. Currently,  metrics are becoming widely
accepted as an easy and balanced way to assist the peer review and
evaluation of scientists and/or research units, provided they have
adequate precision and recall.

This paper presents a benchmarking study of a selected list of
representative Portuguese research units, based on a fairly complete
set of parameters: bibliometric parameters, number of competitive
projects and number of PhDs produced. The study aimed at collecting
productivity and impact data from the selected research units in
comparable conditions i.e., using objective metrics based on public
information, retrievable on-line and/or from official sources and thus
verifiable and repeatable. The study has thus focused on the activity
of the 2003-2006 period, where such data was available from the latest
official evaluation.

The main advantage of our study was the application of automatic
tools, achieving relevant results at a reduced cost. Moreover, the
results over the selected units suggest that this kind of
analyses will be very useful to benchmark scientific
productivity and impact, and assist peer review. 

\end{abstract}
\section{Introduction}
\label{intro}
Bibliometric analysis is becoming widely accepted as an easy and balanced way to measure the research 
impact and relevance of scientists, institutions and even countries~\cite{ball:2005,holden:2006}.
It assumes that citations are references to work that have influenced the author,
and therefore are an evidence of the impact and relevance of the cited work~\cite{macroberts:1989}.
Bibliometric analysis depends mainly on two components:
\begin{description}
\item[Bibliographic Dataset:] from where we retrieve the citations referencing the work of a given scientist;
\item[Citation Metric:] a mathematical formula that produces an unique number quantifying 
                       the impact and relevance of a given scientist from its citations. 
\end{description} 

The most popular bibliographic datasets nowadays are 
Google Scholar\,\footnote{\url{http://scholar.google.com}},
Scopus\,\footnote{\url{http://www.scopus.com/}}
and Web of Science\,\footnote{\url{http:/scientific.thomson.com/isi/}}
(Thomson/Reuters). All have advantages and disadvantages in relation
to the accuracy of data they provide~\cite{nisonger:2004,meho:2007,belew:2005,pauly:2005,bosman:2006},
however, Scopus and Web of Science are subscription-based, which means
that their access is restricted to institutions that subscribe them~\cite{harzing:2007}.  
Furthermore, they only includes citations
published in indexed journals selected by their own
criteria~\cite{roediger:2006}.  Emerging fields such as computer
science and electrical and computer engineering, are particularly
affected by this lack of coverage, as demonstrated by some studies,
notably~\cite{DeSutter:12}.  Thus, although the Web of Science or
Scopus are widely used today, one may question their value for generic
bibliometric analysis, since one would expect this analysis to be
based on a fully-accessible, democratic and comprehensive dataset.  By
contrast, Google Scholar provides a freely available and comprehensive
bibliographic dataset, even if it includes some erroneous entries.

Several citation metrics have been defined and tested, such as the number 
of highly cited papers, the mean number of citations per paper and the total number of citations. 
A recent and popular metric was proposed by Hirsch, the h-index~\cite{hirsch:2005}, defined as follows:
\begin{quote}
A scientist has index $h$ if $h$ of his or her $Np$ papers have at least $h$ citations each and the other $(Np - h)$ 
papers have $\leq$ $h$ citations each. 
\end{quote}
While it has its shortcomings, the appeal of h-index is clear: 
it contributes to the ranking of scientists using 
a single value accounting for production and impact 
that is straightforward to calculate and fairly robust~\cite{ball:2005,glanzel:2005,bornmann:2005,saad:2006}.

In citation metrics, \emph{self-citations} cannot be neglected since they represent 
a significant percentage of the citations~\cite{aksnes:2003,hyland:2003}. 
Self-citation allows authors to connect their recent work to their previously
published findings, and thus are legitimate and necessary to contextualize 
recent work and avoid text repetition.
However, using self-citations for calculating citation metrics would not be reasonable,
since the goal of citation metrics is to measure the scientist's impact on his or her peers. 
Recent works compared different bibliographic datasets~\cite{bosman:2006}, citation metrics~\cite{hirsch:2007}, 
and measured the impact of self-citations~\cite{schreiber:2008}. 

We believe that objective metrics are crucial to evaluate the output
and impact of research units. Lack of completeness, on the one hand,
creates unacceptable competitive disadvantage across research
areas. Lack of precision, like self-citations or miscitations, on the
other hand, compromises the trustworthiness of results.  Attempting to
address these problems, we developed CIDS (Citation Impact
Discerning Self-citations) a tool that automates the post-processing
of raw publication and citation data~\cite{couto2009handling}.  Amongst
other functions, it allows the profiling of publications and citations, both
from individual researchers and whole groups, units or departments.
The root source of data is Google Scholar, which mitigates the completeness
problem. Additionally, the information is post-filtered and cleaned
and in particular, self-citations are removed to address the trustworthiness
problem - a facility we believe is unique in existing tools.
CIDS has been positively evaluated by a number of institutions, both
national and international.

The main advantage of our work is the application of automatic tools
after an initial more labour intensive set-up (e.g. tuning the search
keys).  These tools enable us to extend and update the results with
minimal human intervention and thus achieve relevant results at
reduced cost.  Overall, the results over the selected research units
demonstrate the feasibility of applying such an approach in a research
evaluation setting.  If extended to all units in a given field and
updated regularly, our approach could constitute a very useful tool to
benchmark scientific productivity and impact, and possibly assist the
peer review process.

While bibliometrics are essential to assess research units, they tell
only one part of the story. Looking at the standard practice of
international academic evaluation, we include in this study what we
believe to be a fairly complete set of productivity and impact
metrics: bibliometrics (publications and citations); number of
competitive projects and number of PhDs produced.

Finally, an important facet of trustworthiness is representativity and
reproducibility of the data sets. With that aspect in mind, having a
clear-cut period and set of information is instrumental for the
acceptance of a study by stakeholders and readers.  Such opportunities
are for example given by official research system evaluation cycles,
which provide public information about the aforementioned parameters
of comparable nature in content and period for all units under
evaluation.  Thus, our study focused on research units belonging to
the Portuguese \emph{Fundação para a Ciência e a Tecnologia} (FCT)
sponsored scientific and technological system (``SC\&T'') and was
based on data from latest the FCT evaluation.  The study focused in
particular on research units in our field of interest, the area called
\emph{Engenharia Electrotécnica e Informática (EE\&I)} in the FCT
classification, which encloses what in anglo-saxon terms is described
by the collection of Electrical Engineering and Computer Science and
Engineering. These sectorial benchmarking studies are essential in any
excellence system and common in developed countries. However, they are
not widely disseminated yet in Portugal, so this is our contribution
to that objective.

This paper extends the previous edition~\cite{Couto09a}, along similar
lines with additional research units but the same
reference period.  The objective was to increase representativity of
the sample of selected units, within the limitations of our scarce
team resources, and to significantly revise the structure and
presentation of the study, all in antecipation of the next evaluation
cycle. The paper is structured as follows: Section~\ref{sec:rationale}
introduces the rationale of the study, explaining the reason behind
the parameters and research units selection.  Section~\ref{sec:design}
describes how the study was conducted, explaining its information
sources, information retrieval and processing methodology used, and
the data quality tests performed.  Section~\ref{sec:results} presents
the results obtained in terms of gross and per capita weight and
relevance, and its distribution by unit members.
Section~\ref{sec:conclusions} ends with our main conclusions and
futures directions.

\section{Rationale of the study}
\label{sec:rationale}
We have focused on a specific period, 2003-2006 inclusive, 
since this was the reference period concerning the latest
evaluation\,\footnote{\url{http://alfa.fct.mctes.pt/apoios/unidades/avaliacoes/2007/}}
performed by the Fundação para a Ciência e a Tecnologia (FCT), whose
mission is to continuously promote the advancement of scientific and
technological knowledge in Portugal. The FCT evaluation reported all
units in similar terms so that all units would be in equal conditions,
in respect to information publicly available. Besides the intrinsic
value to our study, a side effect of using the data of the evaluation
period is the ability to match our findings with the very FCT
evaluation results, published in R\&D Units Evaluation Results -
2007\,\footnote{\url{http://alfa.fct.mctes.pt/apoios/unidades/avaliacoes/2007/resultados}}. Regrettably,
that information is only partial for the latest evaluation, since the
results of the evaluation of the research units belonging to associate
laboratories (``LA'', see ahead) were never published.

Fair and open calculation of bibliometric statistics depends on the
availability of a comprehensive database of publications, such as
Google Scholar. To explore Google Scholar we used our freely available
web tool CIDS (Citation Impact Discerning Self-citations) to calculate
bibliometric parameters with and without self-citations. As mentioned
earlier, besides bibliometric, we included other complementary
production parameters in the study: the number of concluded PhDs, and
the competitive national and international projects conducted during
the evaluation period. These parameters complete each other and
together constitute objective indicators of the fulfilment of
qualitative and quantitative goals of a research unit, especially in
comparison with its peers in the same circumstances.

\subsection{Terms of Reference for the units analysed}

In this work, we followed the terms of reference and selection criteria described below.

In the Portuguese S\&T system there are research units and associate
laboratories (``LA''). The latter are larger units, which associate
several formal or informal research units (large groups). LAs are
considered by the government as SC\&T system flagships and are
considerably better financed per PhD than regular units. LAs were part
of the same cycle and reported in the same way as regular units.
Actual timelines have varied according to the real execution of the
process, which involved for example complaints (56\% of the research
units (14/25) complained about the evaluation) and re-evaluations.
Initial evaluations were all based on a 4-year activity report
2003-2006. Re-evaluation results for research units were finally
announced in January 2010, a year later, and 2.5 years after the
evaluation actually started. Evaluation of the LAs was deemed as ended
in 2011, almost 5 years after the activity period concerned, but no
results were made public.

\ED{
Timeline of the EE\&I area (Electrical Engineering and Computer
Science and Engineering) (25 units) evaluation: initial evaluation
started with a 4-year activity report 2003-2006 delivered in July
2007; evaluation results were issued in January 2009; 56\% of the
units (14) complained about the evaluation; re-evaluation results were
announced in January 2010, a year later, and 2.5 years after the
evaluation actually started.
Timeline of the EE\&I area LAs evaluation: initial evaluation started
with the delivery of the 4-year activity report 2003-2006, with some
delays caused by several procedure changes; a first evaluation was
based on the analysis by each LA's Advisory Board; a final evaluation
was issued by a remote ad-hoc panel which performed no visit; it
finally was deemed as ended in 2011, almost 5 years after the activity
period concerned, but no results were made public.
}

We needed a representative set of units for performing our
benchmarking experiments. It was materially impossible to treat all
units, at least in this phase and so, the units were selected to
depict several grades and interesting comparative situations (grading,
initial vs. re-evaluation results, etc.).  Having a mix of stand-alone
research units and LA-based units/groups was also a goal, so we
included three associate laboratories in the study.  ISR and IT are
large LAs composed of several units/groups. We chose ISR Lisbon
(ISR-LX) and the IT unit located in Lisbon (IT-LX).  INESC-ID is a
rather homogeneous LA located in Lisbon.  Overall we selected 8
units, listed in alphabetic order with their main locations: CISTER
(Porto, ISEP), CISUC (Coimbra, FCTUC), CITI (Lisbon, FCTUNL), INESC-ID
(Lisbon, IST), ISRC (Coimbra, FCTUC), ISR-LX (Lisbon, IST), IT-LX
(Lisbon, IST), LaSIGE (Lisbon, FCUL):
\begin{itemize}
\item CISTER, initially rated Very Good (VG), was promoted to Excellent (EX) after re-evaluation.
\item CISUC, initially rated Good (GD), was promoted to VG after re-evaluation.
\item CITI, initially rated GD, remained so after re-evaluation.
\item INESC-ID, the grade was not public at the date of this report.
\item ISRC was the only unit considered Excellent (EX) in the initial
  evaluation.
\item ISR-LX, the grade was not public at the date of this report.
\item IT-LX,  the grade was not public at the date of this report.
\item LaSIGE, initially rated VG, remained so after re-evaluation.
\end{itemize}

We based our experiment on public information, retrievable on-line and/or from
official sources and thus verifiable and repeatable.  Despite our
verifications, the experiment may not be exempt from some residual
errors in individual entries of the source repositories, since it is
based on automated procedures. However, the experiment has a controlled
error margin, as we will discuss subsequently in the Data Quality section.
The error margin is negligible for most of the situations and is similar
across researchers and units. Furthermore, it is better than what could be
achieved by direct query to WoS, GS, DBLNP, Harzing, or related
repositories. Nevertheless, we offered each selected unit the
possibility of verification of their data, but only committed to
correct information which is of official value and obeying the ToR for
the study.

We are primarily interested in producing aggregate data about
institutions, of comparative statistical value. But it should not be
construed from our study that we expect that a simple computation can
be applied to derive an evaluation of a research unit. However,
objective metrics, especially if multi-dimensional and with a good
coverage, are certainly a faithful indicator of the fulfilment
of qualitative and quantitative objectives of a research unit, and
hence an \emph{indispensable tool for peer reviewing within a research
field}.  

This last line prompts for a word of caution about using metrics
directly for comparing productivity and impact of different research
fields, since it is bound to create inacceptable competitive
disadvantages. This is found in some superficial studies and official
bodies' statistics, though it has long been argued to constitute an
unfair practice. Actually, there is now a substantial body of research
scientifically demonstrating these points. Certain indexing methods,
whilst highly competent for classical fields, have drastically lower
precision and recall factors for other, emerging fields, ranging
between 30\% and 60\% lack of coverage in some
cases~\cite{DeSutter:12}. On the other hand, the sheer rate of
production and citation is highly dependent on the field, with e.g.,
average Hirsch-indices of different fields, of researchers of the same
stature and career experience, varying as much as
350\%~\cite{Iglesias:07}.

\ED{This should be intuitively evident if one thinks about factors
  that influence rate of paper production and of development of
  citations, such as: number of practitioners in the field; average
  number of pages and/or authors per paper; average research substance
  per paper; theoretical vs. experimental intensity; etc.}

In summary, we will show below that the parameters chosen for this
study perform well, since they provide a good match to usual
evaluation terms of reference in international academia, including the
official ToRs of the latest FCT evaluation. We hope this will
illustrate the feasibility of applying our methodology and such
parameters in a research unit evaluation setting. We plan on further
extending the study, but the study itself can be extended by anyone
wishing, since the setting and the tools are public.

\section{Study Design} 
\label{sec:design}

A reference parameter of the study is the list of the \emph{unit's
  exclusive integrated researchers with a PhD} \emph{(Int-PhD)}
i.e. its key members, who are not affiliated with another institution.
Int-PhD will be used to: compute aggregate bibliometric indicators;
compute per capita figures of all indicators. We use the Int-PhD list
as of the end of the period in reference (31/12/2006 in this study).

The study focuses on four categories of figures of merit of a unit:
\begin{description}
\item[Weight and Relevance -] measured by the global output and impact
  of the collection of Int-PhD, integrated over a reference
  contributing period.
\item[Production and Impact] - measured by the outputs and impact of
  the unit, specifically over the evaluation period.
\item[Balance] - measured by the distribution of the individual
  Int-PhD's bibliometric figures computed respectively, over the
  reference contributing period, and over the evaluation period.
\item[Efficiency] - measured by the weighting of the above metrics by
  the number of Int-PhDs.
\end{description}

\ED{by their number of publications, number of citations and
  Hirsch-index}

\ED{by their number of publications and number of citations}

The {\bf evaluation period (EP)} in this study is, as explained, the
latest FCT evaluation cycle 4-year period, January 2003 - December
2006 inclusive.

The {\bf reference contributing period (RCP)} is intended to represent the
period of the Int-PhD career's research achievements and experience
that may most directly contribute to the unit. Given that our
objective is the aggregate evaluation of a unit and not of its
individual researchers, we must measure an Int-PhD's contributing
career to the unit and as such, the data about Int-PhD cannot go
arbitrarily back. It has to be in a sufficiently near past considered
to have influenced the current period research, which we have chosen
to be the double of the evaluation period, i.e. an 8-year period from
January 1999 - December 2006 inclusive.

The {\bf balance} metrics are percentile distributions aiming at
characterizing how balanced is the contribution of its key human
resources to the relevance (long-term indicators) and impact
(short-term indicators) metrics.

We compute {\em gross} and {\em per capita} metrics, since it is
fundamental to distinguish between the critical mass of a unit, and
the {\bf efficiency} with which it puts that critical mass at work.
In concrete terms, this amounts to making the difference between
\emph{production} of a collection of Int-PhD researchers, e.g. in
number of papers or theses, and \emph{productivity} of that
collection, e.g. in number of papers or theses per Int-PhD researcher
(or per euro of financing, for that matter). Other figures of merit
notwithstanding, efficiency is becoming a primal figure of merit
to assess the return of financing of research units in comparable
conditions.

\subsection{Information Sources} 

The idea was to gather a number of parameters that could be
automatically calculated and would be sufficient to derive an
evaluation of a research unit, in terms of the three categories of
figures of merit introduced above. 

In order to guarantee the fairness and repeatability of the study, we
postulated the following rules for the parameters:
\begin{itemize}
\item be based on a known and generic formula and thus repeatable;
\item be applicable to every unit;
\item be based on public information, retrievable on-line and/or from
  official sources and thus verifiable and reproducible.
\end{itemize}

Besides bibliometric parameters, we included two other measurable
output items that satisfied the above rules: the number of concluded
PhDs, and the national and international projects conducted during the
evaluation period. Overall we selected and computed the following
parameters:
\begin{itemize}

\item {\bf Weight and Relevance (gross)}
\begin{enumerate}\small
\item Number of Int-PhD at the end of the evaluation period
\item Number of unique cited papers over the reference contributing period 
\item Number of unique citations to papers published over the reference contributing period
\end{enumerate}

\item {\bf Production and Impact (gross)}
\begin{enumerate}\small
\item Number of unique cited papers over the evaluation period
\item Number of unique citations to papers published over the evaluation period
\item Number of international and national competitive research projects started during the evaluation period
\item Number of PhD theses produced during the evaluation period
\end{enumerate}

\item {\bf Efficiency - Weight and Relevance (per capita)}
\begin{enumerate}\small
\item Number of unique cited papers for each Int-PhD over the reference contributing period 
\item Number of unique citations per Int-PhD over the reference contributing period 
\item Average Hirsch-index of Int-PhDs over the reference contributing period
\end{enumerate}

\item {\bf Efficiency - Production and Impact (per capita)}
\begin{enumerate}\small
\item Number of unique cited papers for each Int-PhD over the evaluation period
\item Number of unique citations per Int-PhD to papers published over the evaluation period
\item Number of international and national competitive research projects per Int-PhD started during the evaluation period
\item Number of PhD theses produced per Int-PhD during the evaluation period
\end{enumerate}

\item {\bf Balance - Relevance}
\begin{enumerate}\small
\item Distribution of the Int-PhD's numbers of cited papers over the reference contributing period 
\item Distribution of the Int-PhD's numbers of citations over the reference contributing period
\item Distribution of the Int-PhD's Hirsch-index over the reference contributing period
\end{enumerate}

\item {\bf Balance - Impact}
\begin{enumerate}\small
\item Distribution of the Int-PhD's numbers of cited papers over the evaluation period 
\item Distribution of the Int-PhD's numbers of citations over the evaluation period
\end{enumerate}
\end{itemize}

The number of unique papers and citations represents the union of the
set of papers and citations found for each individual Integrated PhD
researcher, thus eliminating repetitions. For example, papers
co-authored by unit researchers are only counted once.

As a note, these metrics cover well the several quantitative
aspects normally at stake by international criteria, when evaluating a
research unit or group or department. Incidentally, they also end-up
representing well the quantitatively measurable aspects of the
FCT evaluation philosophy, at least judging from the ToR for the
latest evaluation:
\begin{itemize}
\item Productivity (papers) 
\item Relevance/Impact (citations, h-index ) 
\item Feasibility (projects) 
\item Training (PhDs theses) 
\end{itemize}

Thus, our study may provide some insight on
the FCT unit's evaluation results vs. criteria.

\subsection{Information retrieval and processing methodology}

The target data of this study was thus:
\begin{itemize}
\item The publications and citations of Int-PhD measured over two
  periods: reference contributing period (99-06); and the evaluation
  period (03-06).
\item The PhD theses and projects of each unit measured over the
  evaluation period (03-06).
\end{itemize}

The calculation of the parameters was based on the following sources of information:
\begin{itemize}
\item Google Scholar (GS) repository (corrected, post-processed and
  filtered by the CIDS tool).
\item FCT web site.
\item Multi-annual evaluation report 2003-2006 from units (to the
  exception of ISRC, whose report was not made available to us;
  nevertheless, the missing unit's data was retrieved from the unit's
  and FCT's site).
\item Units' web sites.
\end{itemize}

Our first step was to obtain the list of Int-PhD researchers of each
unit at 31/12/2006 from the FCT web site. From the FCT web site we
could not collect the list of Int-PhD researchers for older dates. For
each researcher, we manually defined a Google Scholar query that best
defined his/her list of published papers. This list of queries was
given as input to our tool CIDS, a freely available tool that
automatically calculates bibliometric parameters based on Google
Scholar data. Given the importance of bibliometric parameters in our
study, we provide a detailed description of CIDS in a following
section. The queries for the researchers were manually updated and executed 
in 2012. 

The number of national and international projects, and the number of
concluded PhD theses were collected from the unit's evaluation
reports, cross-checked with the unit's web site or other official
sites when needed. We had access to all unit's evaluation reports to
the exception of ISRC, whose report was not made available to us;
nevertheless, the missing unit's data was retrieved from the unit's
and FCT's site.

\subsection{CIDS}
To calculate the citation metrics for each selected author, the
current version of CIDS\,\footnote{\url{http://cids.fc.ul.pt}} only
requires a Google Scholar query, normally the last name of the author
together with his/her initials\,\footnote{Previous releases of CIDS
  featured the subject area \emph{(subject:)} operator, which is no
  longer supported by GS.}. Besides the \emph{(author:)} operator,
any other of the Advanced Scholar Search
operators can be included\,\footnote{\url{http://scholar.google.pt/advanced_scholar_search}}.

The papers returned by Google Scholar are then individually analyzed. 
For each paper, CIDS retrieves its citations and its authors' names. 
CIDS uses the authors' names to filter out the self-citations 
based on the self-citation policy of CiteSeer~\cite{lawrence1999digital}. 
CIDS current policy is: \emph{marking a citation as a self-citation if at least one of its 
authors is also an author of the cited paper}.
In the end, CIDS uses the number citations
of each paper to calculate the h-index, the citations-per-paper,
and the total number of citations, and uses the number of non-self-citations 
to calculate the same citation metrics. Thus, CIDS returns two values for each 
citation metric, one using all citations and the other discerning self-citations. 

\begin{figure*}
\resizebox{\textwidth}{!}{%
  \includegraphics{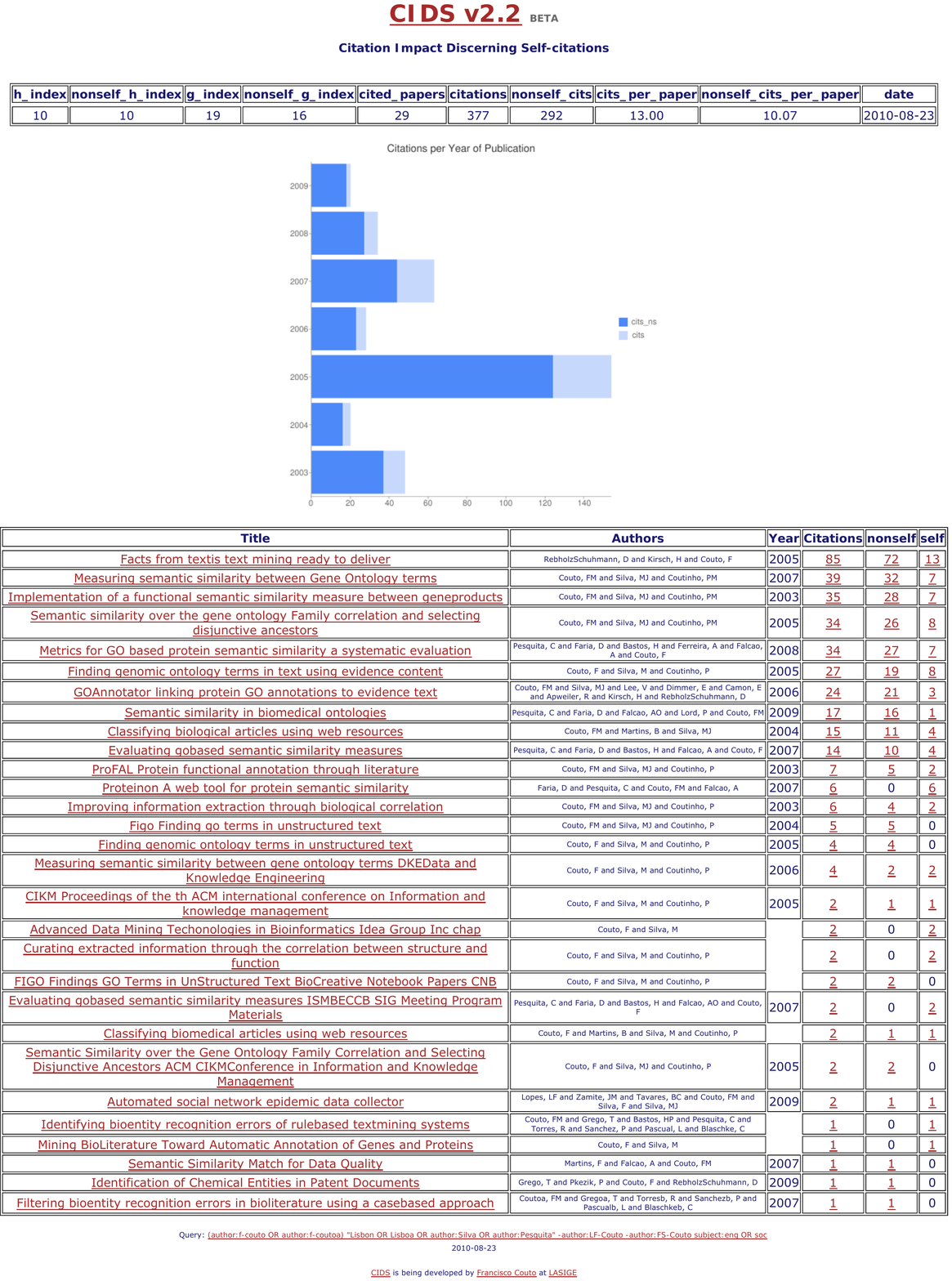}
}
\caption{Example of an individual's bibliometric analysis provided by CIDS}
\label{fig:cids_individual}
\end{figure*}

For example, the query  producing the results shown in Figure~\ref{fig:cids_individual},
used \emph{'Lisbon OR Lisboa' -author:LF-Couto} to disambiguate the author's name,
by only selecting authors from Lisbon and discarding the author with the 
initials LF\,\footnote{\url{http://scholar.google.com/intl/en/scholar/refinesearch.html}}.
The first table shows the values for each citation metric with and without including self-citations.
The second table shows the number of citations, the number of self-citations, and the number of
non-self-citations. Each number is a link to obtain the respective list of citations.
Besides HTML, the tool also provides the citation analysis in TSV and BibTeX formats. 

A list of individuals can be assigned to a research unit
to produce aggregate values.
CIDS calculates two groups of aggregate values: the unique values and the average values.
Unique values are calculated by merging the papers and citations 
found for all individuals. Thus, these unique values 
just consider a paper or a citation once, even if it is shared 
by multiple individuals from the same research unit.
Average results are calculated just by averaging the individual values
for each bibliographic metric.

\subsection{Data Quality}
The accuracy of CIDS depends on the ability of Google Scholar's method to correctly identify the names of the authors in the header of the paper. The method is robust in general, since it is relatively simple to automatically detect the header of a paper, with a small error margin. However, a few authors have ambiguous names that can lead CIDS to include papers from homonymous authors~\cite{kang2009co}. The impact of this problem in our study is residual, and since we aim at evaluating a group of researchers and not specific individuals, we can consider it negligible. However, in order to eliminate any outlier in this particular study, each query was manually verified.

For evaluating the accuracy of CIDS, we crosschecked a manually curated list of 129 cited papers of an Int-PhD researcher with the papers automatically identified by CIDS. We found that 103 of the 105 papers returned by CIDS were in the curated list. This means that CIDS achieved a precision of 96\% and a recall of 78\%. Moreover, the real recall of CIDS is expected to be even higher than 78\%, since in our study CIDS was limited to the first two Scholar result pages for each query due to performance issues, and senior researchers (as was the case) tend to pass this limit.

Considering the existence of other public and well-organised repositories, we made a comparative study of the precision and recall with DBLP, another reference repository. We crosschecked the same manual list with the list of papers assigned by DBLP. We found that 90 of the 91 papers returned by DBLP were also in the curated list. This gives a precision of 99\% for DBLP but a recall of only 70\%. We also found that all the papers in DBLP were also available in Scholar, which means that, barring one or another exception, including DBLP will not represent an improvement on recall. 

We stress that using our tool for individual purposes (e.g. a curriculum) will require a final albeit residual effort of checking and cleaning. That effort seems minimal, as reported by the additional experiment below. We compared the manually curated list of papers and citations of another Int-PhD researcher with the results returned by CIDS. The curated list contained 69 papers and 211 nonself-citations, whereas CIDS returned 67 papers and 207 nonself-citations. Since all the papers and nonself-citations returned by CIDS were also in the cleaned list, we obtained a precision of 100\% and a recall of 97\% for papers and 98\% for non-self citations. 
This demonstrates that our results based on Scholar queries are quite accurate and complete. 

Another issue with Google Scholar (and in general with any automated
tool) is the duplication of data, as the same paper can appear multiple
times in different entries. This issue influences the number of cited
papers and possibly h-index parameters, but not the total citation count.
To evaluate the real impact of this issue we calculated the number of
distinct Scholar entry pairs with equal titles. We found only 68 pairs
from 4,532 distinct entries, which means that the issue affects less
than 1.5\% of the entries. Furthermore, since most citations tend to
be assigned to a single entry in the cases of duplication, the h-index
will normally not be affected.


\section{Results}
\label{sec:results}

\subsection{Gross Weight and Relevance (gross)} 
Gross results are useful to measure the critical mass of the unit,
based on the global weight and relevance of the collection of its
Integrated PhD researchers, over their contributing career to the
unit. However, they are also biased by the seniority and the size
of the unit, as units with more researchers and in particular with
more senior researchers will tend to to accumulate more papers
and citations. Thus, they do not account for a unit's efficiency
and effectiveness which we will discuss subsequently.

\begin{figure*}
\resizebox{\textwidth}{!}{%
  \includegraphics{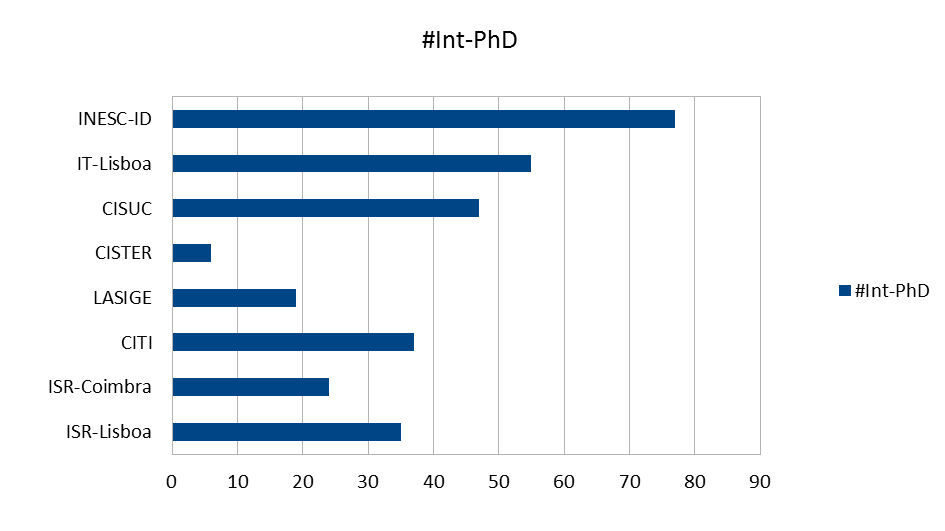}
}
\caption{Number of Exclusive Integrated PhD researchers of the unit
  (\#Int-PhD) at the end of the evaluation period (EP).}
\label{fig:int-phd}
\end{figure*}

\begin{figure*}
\resizebox{\textwidth}{!}{%
  \includegraphics{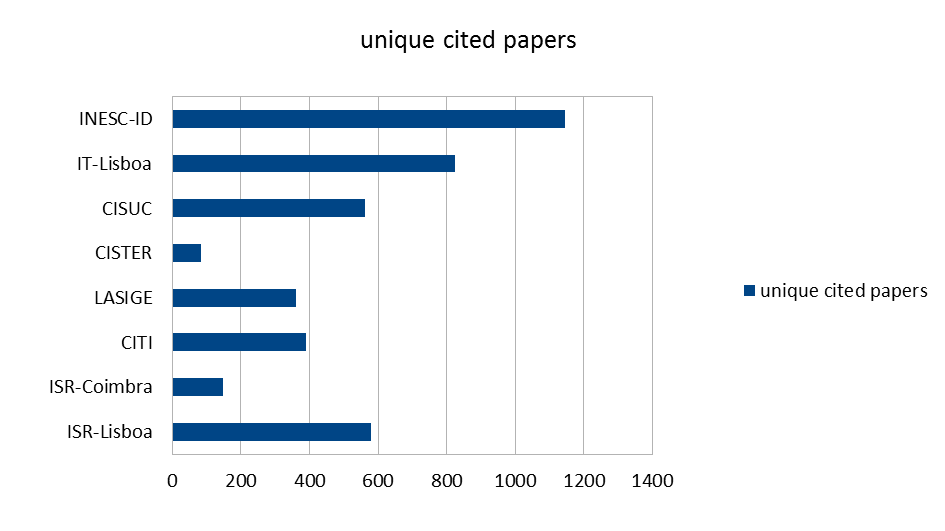}
}
\caption{Unique cited papers: union of the sets of papers found for each individual Int-PhD 
published within the reference contributing period, RCP (1999-2006).}
\label{fig:unique-cited-papers}
\end{figure*}

\begin{figure*}
\resizebox{\textwidth}{!}{%
  \includegraphics{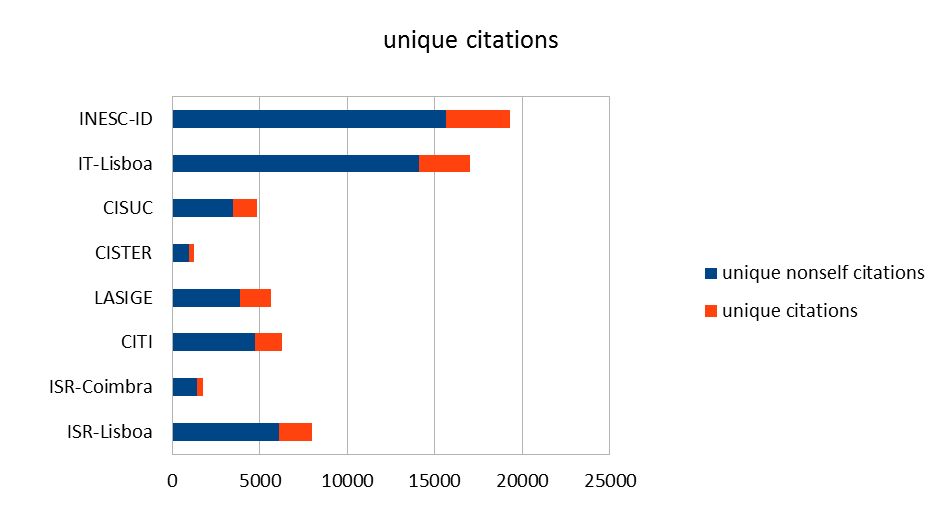}
}
\caption{Unique citations: union of the sets of citations found to
  each of the papers from each individual Int-PhD published within the
  RCP (1999-2006).}
\label{fig:unique-citations}
\end{figure*}

Gross results that were calculated over the reference contributing
period (99-06):
\begin{enumerate}
\item Number of exclusive integrated PhD researchers of the unit at
  the end of the evaluation period (\#Int-PhD) (Figure
  \ref{fig:int-phd}).
\item Number of unique cited papers (Figure
  \ref{fig:unique-cited-papers}): global publication figure created
  from the union of the papers found (with at least one citation) over
  the reference contributing period, from each individual Integrated
  PhD researcher (thus eliminating repetitions, e.g., papers
  co-authored by unit researchers are only referred once).
\item Unique citations (Figure \ref{fig:unique-citations}): global
  citation figures created from the union of citations found to each
  of the papers calculated above (thus eliminating repetitions, e.g.,
  citations to papers co-authored by unit researchers are only
  referred once).
\end{enumerate}

\subsection{Production and Impact (gross)}
Outputs over a period of time provide a measure of the unit's
effectiveness with regard to production (publications, research
projects and PhD theses) and corresponding impact (citations). Instead
of measuring what a unit seems capable of doing (weight and relevance)
they measure what a unit has actually done in a given period of time.
However, gross production and impact results are still biased by the
size of the unit, as units with more researchers tend to produce more
papers and citations per period of time. Thus, these metrics also
do not account for a unit's efficiency, and are of limited use for
comparing research units that differ greatly in size.

\ED{Since there is an evident lag between publication and citations,
it only makes sense to evaluate the impact of given production period
after a time interval to account for that lag. Thus, we propose to
evaluate the output of an $n$ year production period, at
least $n$ years later. That is, for the 4-year evaluation period
03-06, we will evaluate citations made at least by 2010. 
}

\begin{figure*}
\resizebox{\textwidth}{!}{%
  \includegraphics{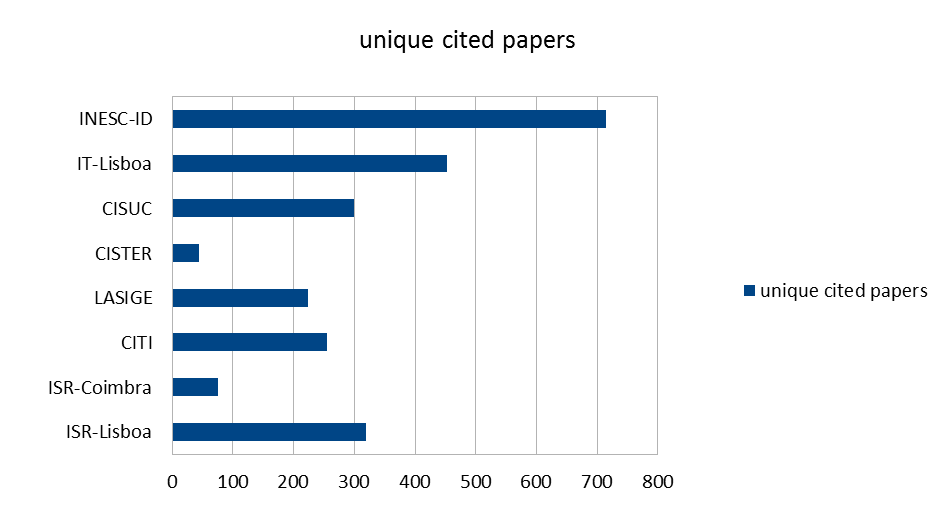}
}
\caption{Unique cited papers: union of the sets of papers found for
  each individual Int-PhD and published within the evaluation period,
  EP (2003-2006).}
\label{fig:unique-cited-papers-0306}       
\end{figure*}

\begin{figure*}
\resizebox{\textwidth}{!}{%
  \includegraphics{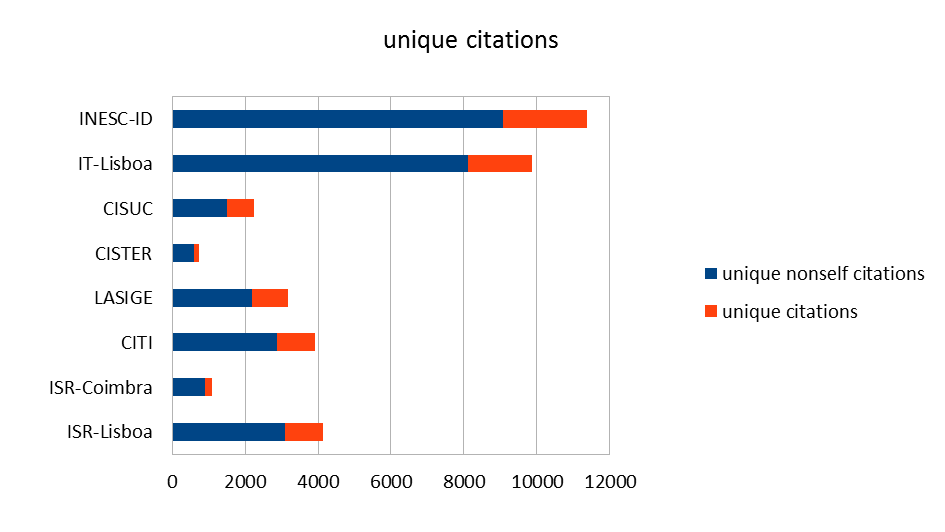}
}
\caption{Unique citations: union of the sets of citations found to
  each of the papers from each individual Int-PhD published within the
  EP (2003-2006).}
\label{fig:unique-citations-0306}       
\end{figure*}

\begin{figure*}
\resizebox{\textwidth}{!}{%
  \includegraphics{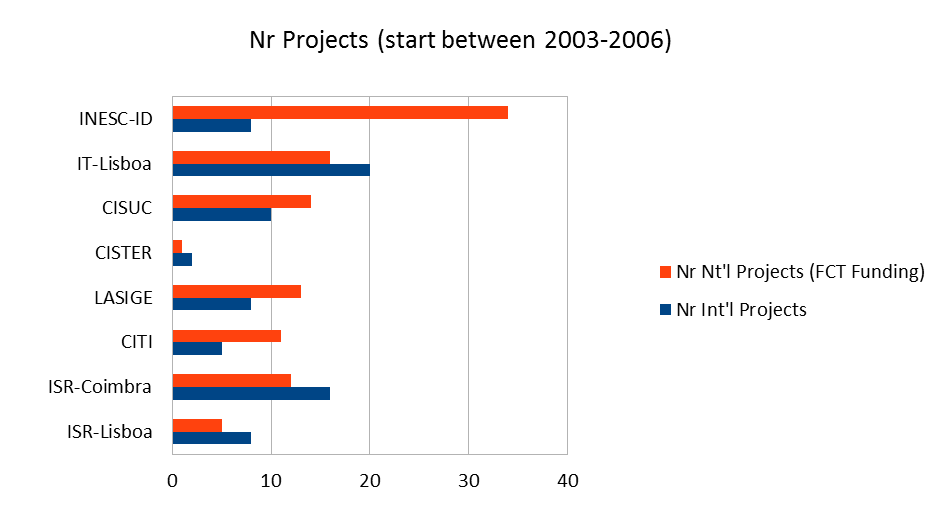}
}
\caption{National and International projects: numbers of research projects started during the
  EP (2003-2006).}
\label{fig:projects}       
\end{figure*}

\begin{figure*}
\resizebox{\textwidth}{!}{%
  \includegraphics{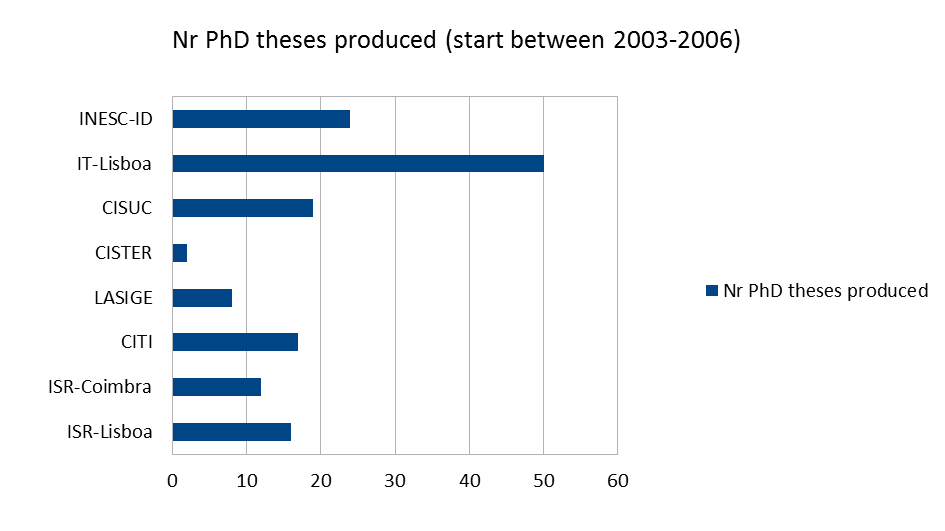}
}
\caption{PhD theses produced during the EP (2003-2006).}
\label{fig:theses}       
\end{figure*}

Gross results that were calculated over the evaluation period (03-06):
\begin{enumerate}
\item Unique cited papers (Figure \ref{fig:unique-cited-papers-0306}): union of the papers found from each individual Int-PhD published in the period. 
\item Unique citations (Figure \ref{fig:unique-citations-0306}): union of citations found to each of those papers. 
\item National and International projects (Figure \ref{fig:projects}): numbers of research projects started during the period. 
\item PhD theses produced (Figure \ref{fig:theses}): numbers of PhD theses finished during the period. 
\end{enumerate}

\subsection{Efficiency - Weight and Relevance per Capita}
Weight and relevance per capita results (e.g., figures 'per InT-PhD')
provide some measure of a unit's relative density, by
dividing the gross publication and citation figures (over the reference
contributing period) by the number of Int-PhD.

These metrics enable us to compare units directly, irrespective
of their size, since they measure the unit's normalized critical mass.
Special emphasis should be given to the h-index, a true measure of substance and consistency of both
production and impact \emph{over the years}, since an author's h-index
is given by the highest number $n$ of papers with at least $n$
citations.

\begin{figure*}
\resizebox{\textwidth}{!}{%
  \includegraphics{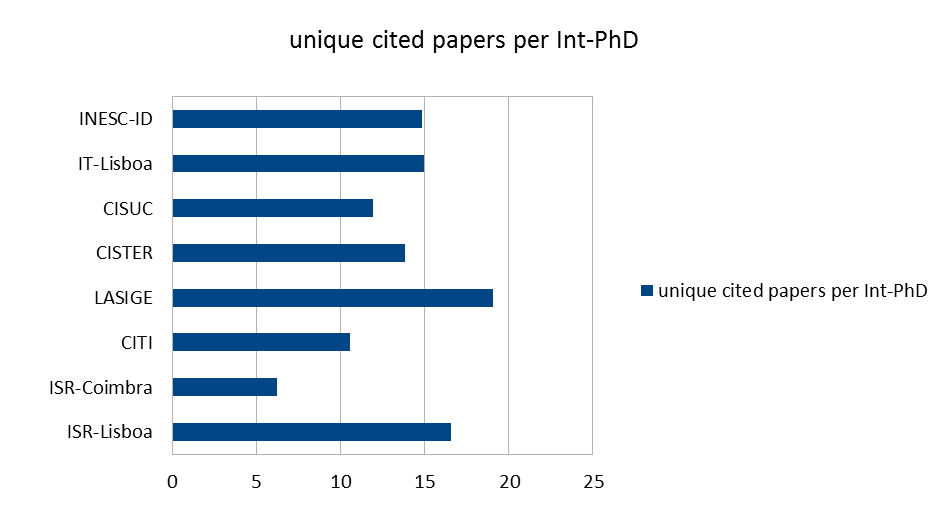}
}
\caption{Unique cited papers per Int-PhD: Gross Weight publication figure divided by \#Int-PhD w.r.t. the
  RCP (1999-2006).}
\label{fig:unique-cited-papers-per-phd}       
\end{figure*}

\begin{figure*}
\resizebox{\textwidth}{!}{%
  \includegraphics{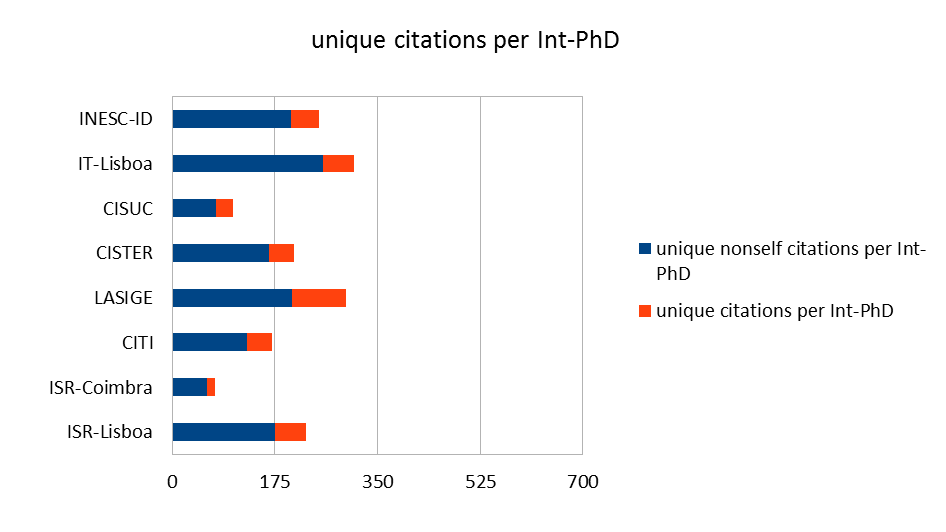}
}
\caption{Unique citations per Int-PhD: Gross Weight citations figure
  divided by \#Int-PhD, w.r.t. the RCP (1999-2006).}
\label{fig:unique-citations-per-phd}       
\end{figure*}

\begin{figure*}
\resizebox{\textwidth}{!}{%
  \includegraphics{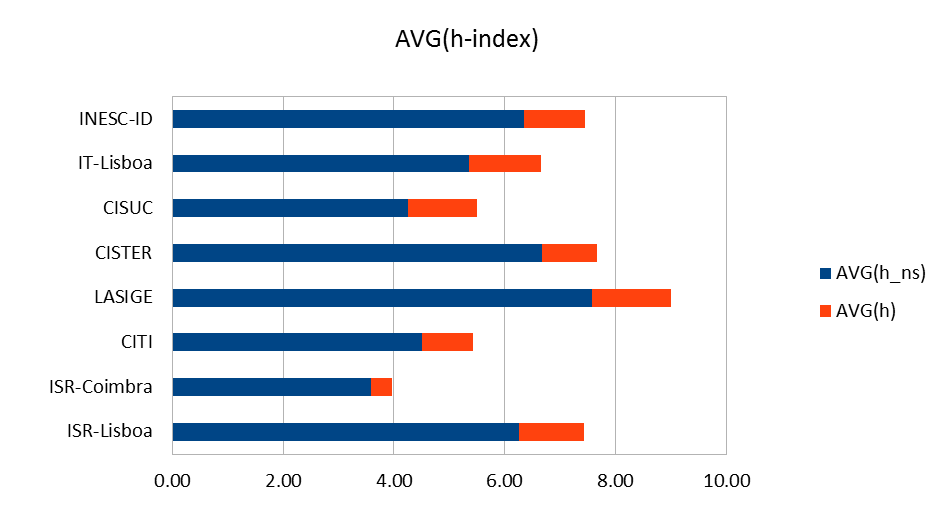}
}
\caption{Average of the h-indices of Int-PhD, w.r.t. the RCP (1999-2006).}
\label{fig:avg-h-index}       
\end{figure*}

Weight and relevance results per capita, calculated over the reference contributing
period (99-06):
\begin{enumerate}
\item Unique cited papers per Int-PhD (Figure
  \ref{fig:unique-cited-papers-per-phd}): Gross publication figure
  divided by \#Int-PhD.
\item Unique citations per Int-PhD (Figure
  \ref{fig:unique-citations-per-phd}): Gross citations figure divided
  by \#Int-PhD.
\item Average of the h-indices (Figure \ref{fig:avg-h-index}): sum of the h-index of each Int-PhD, divided by \#Int-PhD.
\end{enumerate}

\subsection{Efficiency - Production and Impact per capita}
While the gross outputs over a period of time measure a unit's effectiveness,
it is also important to assess its efficiency with regard to production
(publications, research projects and PhD theses) and respective impact
(citations). This was done by dividing the gross production and
impact figures for the same evaluation period by the number of Int-PhD.
Production and impact per capita (e.g., figures 'per Int-PhD') are
the most suitable metrics to compare research units because they are
not affected by the number or seniority of researchers, but rather
reflect the average productivity and impact of the researchers in a unit.

\begin{figure*}
\resizebox{\textwidth}{!}{%
  \includegraphics{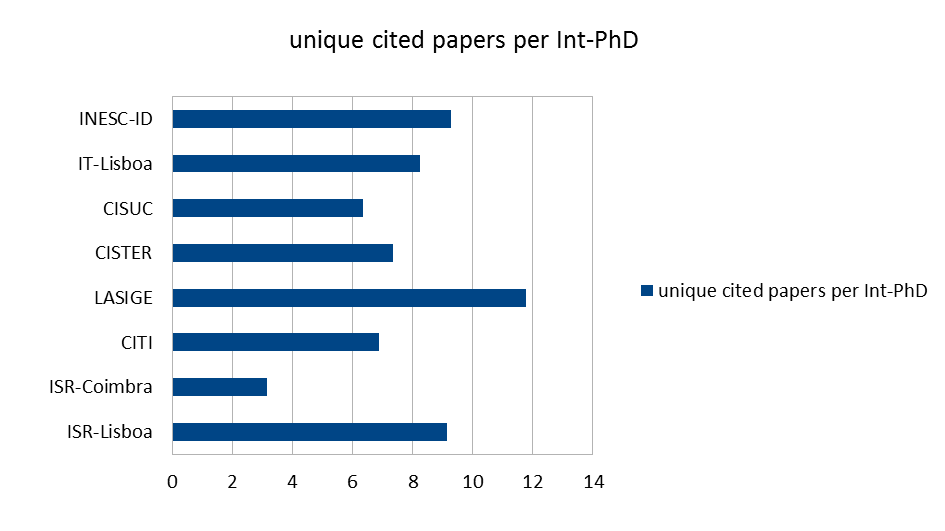}
}
\caption{Unique cited papers per Int-PhD: Gross Weight publication figure divided by \#Int-PhD w.r.t. the
  EP (2003-2006).}
\label{fig:unique-cited-papers-per-phd-0306}
\end{figure*}

\begin{figure*}
\resizebox{\textwidth}{!}{%
  \includegraphics{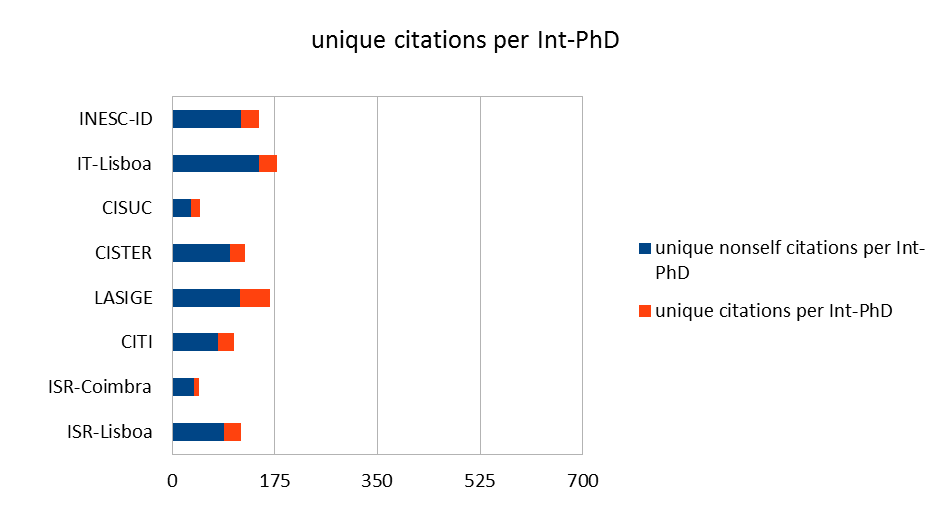}
}
\caption{Unique citations per Int-PhD: Gross Weight citations figure
  divided by \#Int-PhD w.r.t. the EP (2003-2006).}
\label{fig:unique-citations-per-phd-0306}
\end{figure*}

\begin{figure*}
\resizebox{\textwidth}{!}{%
  \includegraphics{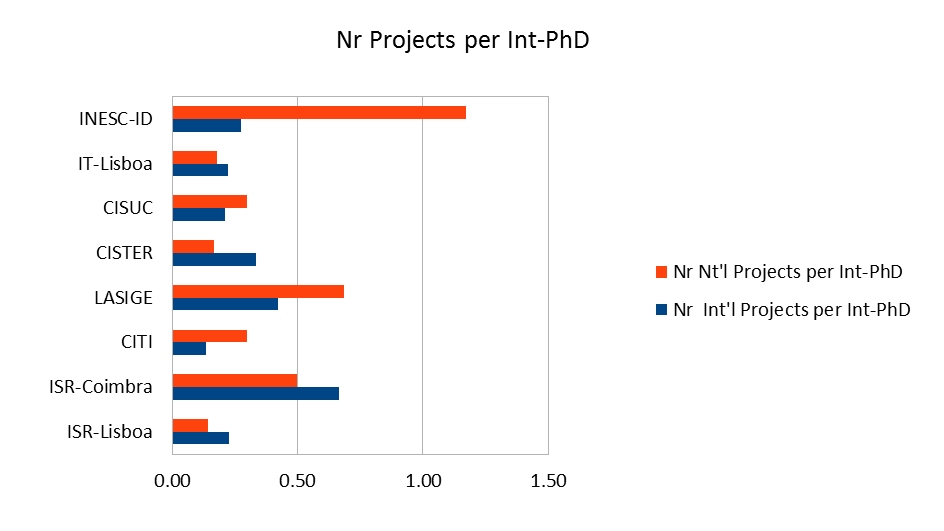}
}
\caption{National and International projects per Int-PhD started
  during the EP (2003-2006): Gross Weight research projects figure
  divided by \#Int-PhD.}
\label{fig:projects-per-phd}
\end{figure*}

\begin{figure*}
\resizebox{\textwidth}{!}{%
  \includegraphics{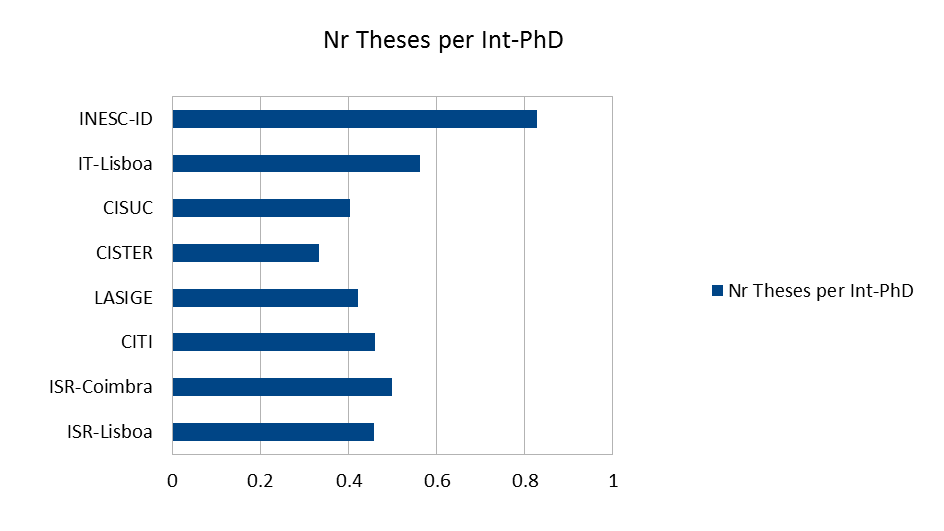}
}
\caption{PhD theses produced per Int-PhD produced during the EP
  (2003-2006): Gross Weight PhD theses figure divided by \#Int-PhD.}
\label{fig:theses-per-phd}       
\end{figure*}

Production and impact results per capita, calculated over the evaluation period (03-06):
\begin{enumerate}
\item Unique cited papers per Int-PhD (Figure
  \ref{fig:unique-cited-papers-per-phd-0306}): union of the papers
  published in the period, found from each individual Int-PhD, divided
  by \#Int-PhD.
\item Unique citations per Int-PhD (Figure
  \ref{fig:unique-citations-per-phd-0306}): union of citations found
  to each of those papers, divided by \#Int-PhD.
\item National and International projects per \#Int-PhD
  (Figure \ref{fig:projects-per-phd}): numbers of research projects
  started during the period, divided by $10\times\#Int-PhD$ (for
  readability).
\item PhD theses produced per \#Int-PhD (Figure
  \ref{fig:theses-per-phd}): numbers of PhD theses finished during the
  period, divided by $10\times\#Int-PhD$ (for readability).
\end{enumerate}

\subsection{Balance - Relevance}
These metrics estimate the distribution of the relevance of individual
Int-PhD unit members, for each unit. They enable the comparison of research
units regardless of their size, since the distribution is relative to the
number of Int-PhDs.

The function QNT(\emph{parameter}) measures the percentage of Int-PhDs
of each unit that fall between selected threshold values
of \emph{parameter}. For example, \% of researchers with: up
to 50 papers; 51-100; 101-150; above 150.

\begin{figure*}
\resizebox{\textwidth}{!}{%
  \includegraphics{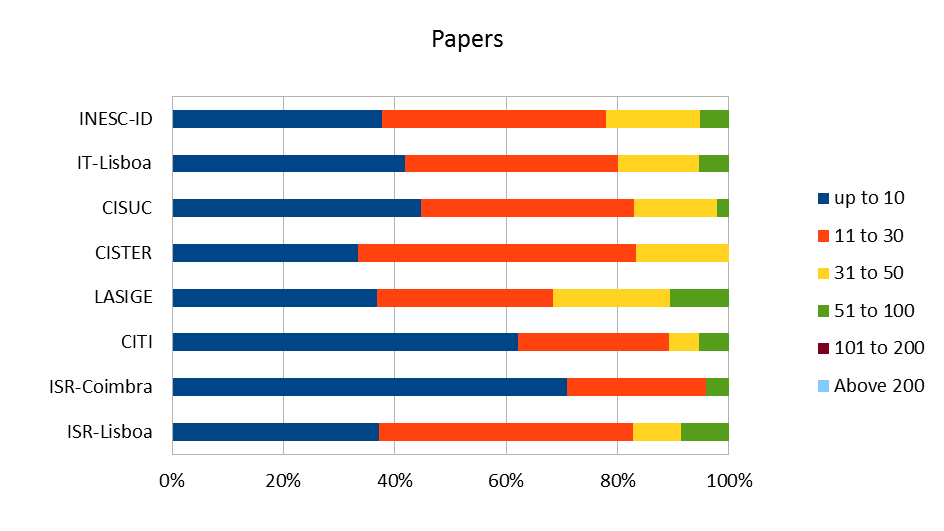}
}
\caption{Distribution of Int-PhD researchers by their number of cited papers published within the
  RCP (1999-2006).}
\label{fig:distribution-papers}
\end{figure*}

\begin{figure*}
\resizebox{\textwidth}{!}{%
  \includegraphics{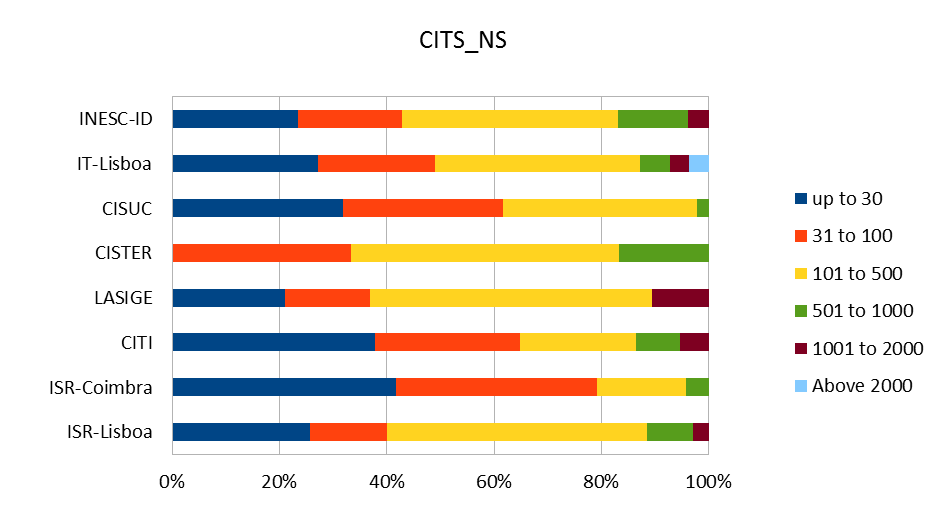}
}
\caption{Distribution of Int-PhD researchers by their number of citations (excluding self-citations) to the papers published within the
  RCP (1999-2006).}
\label{fig:distribution-citations}
\end{figure*}

\begin{figure*}
\resizebox{\textwidth}{!}{%
  \includegraphics{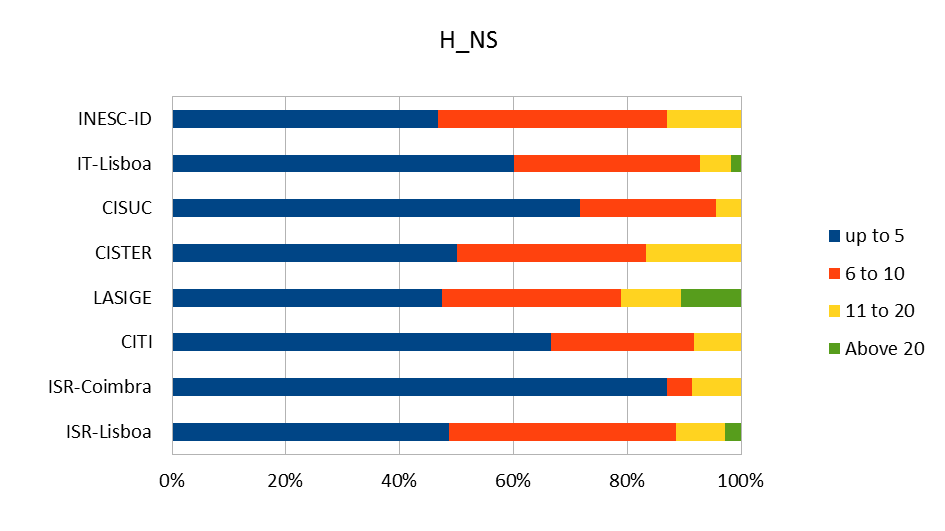}
}
\caption{Distribution of Int-PhD researchers by their h-index
  (excluding self-citations) w.r.t. publications within the RCP
  (1999-2006).}
\label{fig:distribution-h-index}       
\end{figure*}

Results are shown for the distribution of the number of
cited papers, citations and Hirsch-index, excluding self-citations
in the latter two. Together, they yield a macroscopic estimate of how
balanced each unit is in terms of relevance of its members. The larger
the rightmost bars are in the figures \ref{fig:distribution-papers}, \ref{fig:distribution-citations} and \ref{fig:distribution-h-index}, 
the better balanced is each unit. Again, special attention should be drawn to the h-index
distributions.

Distributions (QNT (papers | CITS NS | H NS)) that were calculated
over the reference contributing period (99-06):
\begin{enumerate}
\item Distribution of the Int-PhD's numbers of cited papers
  (Figure~\ref{fig:distribution-papers}).
\item Distribution of the Int-PhD's numbers of citations (excluding self-citations)
  (Figure~\ref{fig:distribution-citations}).
\item Distribution of the Int-PhD's Hirsch-index (excluding self-citations)
  (Figure~\ref{fig:distribution-h-index}).
\end{enumerate}

\subsection{Balance - Impact}

These metrics estimate the distribution of the impact of individual
Int-PhD unit members, for each unit. Like the relevance metrics,
they enable the direct comparison of research units regardless of their size,
since the distribution is relative to the number of Int-PhDs.
Again, we are using the function QNT(\emph{parameter}) as defined in the previous
section.

\begin{figure*}
\resizebox{\textwidth}{!}{%
  \includegraphics{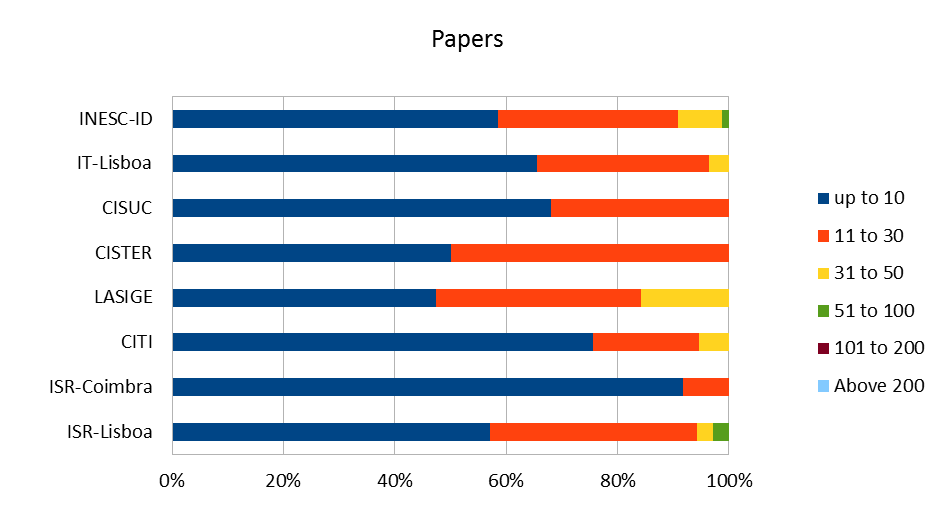}
}
\caption{Distribution of Int-PhD researchers by their number of cited papers published within the
  EP (2003-2006).}
\label{fig:distribution-papers-0306}       
\end{figure*}

\begin{figure*}
\resizebox{\textwidth}{!}{%
  \includegraphics{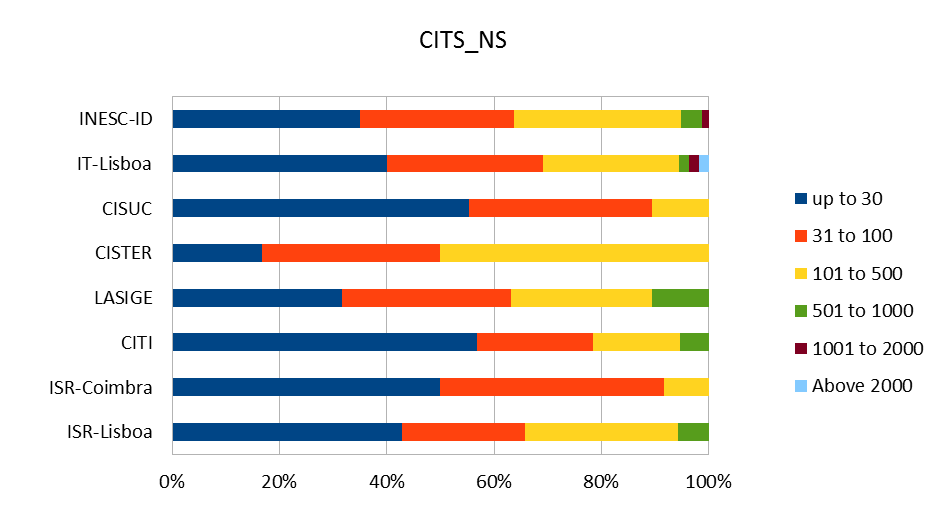}
}
\caption{Distribution of Int-PhD researchers by their number of
  citations (excluding self-citations) to the papers published within
  the EP (2003-2006).}
\label{fig:distribution-citations-0306}       
\end{figure*}

Results are shown for the distribution of number of
cited papers published in the evaluation period, and their citations
excluding self-citations. Together, they yield a macroscopic
estimate of how balanced each unit has been, in terms of the
contributions of individual Int-PhD researchers to its impact over a
period. Again, the larger the rightmost bars are in the Figures 
\ref{fig:distribution-papers-0306} and \ref{fig:distribution-citations-0306}, 
the better balanced is each unit.

As explained previously, for the 4-year period
03-06, we are evaluating citations more than four years later. Note that
h-index is not included since it does not apply to short periods.

Distributions (QNT (papers | CITS NS)) that were calculated
over the evaluation period (03-06):
\begin{enumerate}
\item Distribution of the Int-PhD's numbers of cited papers
  (Figure~\ref{fig:distribution-papers-0306}).
\item Distribution of the Int-PhD's numbers of citations (excluding self-citations)
  (Figure~\ref{fig:distribution-citations-0306}).
\end{enumerate}

\section{Conclusions}
\label{sec:conclusions}
This paper presented a study that compared a set of representative
Portuguese research units using objective parameters. The calculations
of these parameters were based on public information, retrievable
on-line and/or from official sources and thus verifiable and
repeatable.  
The results have shown that the parameters chosen for this study
perform well, since they allowed to produce aggregate data about
institutions, of comparative statistical value, providing a good match
to usual evaluation terms of reference in international academia,
including the official ToRs of the latest FCT evaluation.

This kind of benchmarking studies are essential in any excellence
system, and common in developed countries, but they are normally
expensive and specific to a given period and domain. By contrast,
our study required minimal human intervention, since it collected most
of the information using automatic tools, such as CIDS, from publicly
available resources. This resulted in the analysis of a set of
extensive information that can be easily kept up to date, since we can
track public data sources automatically for updates as they evolve.
Moreover, our approach could be easily extended to other fields as
long as similar sources of information are available.  We plan on
extending the present study, but it can also be extended by
anyone willing.

The main goal of this study was to calculate and show objective
numbers, avoiding controversial discussions about the chosen
parameters.  However, in the future we plan to perform more extensive
sensitivity analyses, for example, to verify the effect of discerning
self-citations and to measure the impact of homonymous authors~\cite{galhardas2011support}.  
For doing this, we will look to available datasets containing
manually verified associations of publications and citations to
authors. We also plan on evolving the CIDS tool itself to improve its
efficiency and accuracy. One avenue that is being explored
in a beta version of a new release of CIDS is to take
advantage of the Google Scholar Citation profiles, which requires the
collaboration of the target units and researchers.
We stress that CIDS can and has been used for individual purposes
(e.g., a curriculum) but we recommend a final albeit residual effort
of checking and cleaning. That effort is predicted to be minimal, as
reported by the experiments on data quality. 

\ED{This will become even easier with the next version
of CIDS that will use AJAX technology to provide a preview of the
results, which can be verified as they are completed on-the-fly with
the missing numbers.}

\ED{In a soon to be released
version of CIDS, this effort may be performed right at the source, in
Google Scholar Citations, CIDS still remaining a valuable tool for
discerning self-citations and for performing aggregate unit studies.}

Finally, the results over the selected units suggest that
objective metrics, especially if multi-dimensional and with
good precision and recall, are a faithful indicator of the
fulfilment of qualitative and quantitative objectives
of a research unit. As such, they can be a useful tool to benchmark
scientific productivity and impact, and assist peer review.

\section*{Acknowledgements}
We would like to thank Ana Luisa Respício for the valuable advice on
statistics; Pedro Antunes for the manual evaluation of CIDS results;
Helena Galhardas and Emanuel Santos for calculating the number of
Scholar duplicates; Luís Caires, Luís Rodrigues, Mário Silva, and
Eduardo Tovar, for their many valuable suggestions and comments, and
several other researchers for point suggestions and for encouraging us
to pursue this avenue. We would like to thank all LaSIGE members who
used the tool and reviewed the results, and Ivan Andrade for his
contribution to the first edition, whose results are extensively re-used.

%
 \bibliographystyle{plain}
 \bibliography{cids}
%
%
%

\end{document}